\documentclass[prl,twocolumn,showpacs,preprintnumbers,amsmath,amssymb]{revtex4}
\newcommand{\ket}[1]{\left | #1 \right\rangle}

\usepackage[dvips]{graphics}
\usepackage{tikz}

\usepackage{amsmath}

\DeclareMathOperator{\arcsinh}{arcsinh}

\begin{document}

\title{On the testability of the equivalence principle as a gauge principle detecting the gravitational $t^3$ phase}

\author{Chiara Marletto and Vlatko Vedral}
\affiliation{Clarendon Laboratory, University of Oxford, Parks Road, Oxford OX1 3PU, United Kingdom and\\Centre for Quantum Technologies, National University of Singapore, 3 Science Drive 2, Singapore 117543 and\\
Department of Physics, National University of Singapore, 2 Science Drive 3, Singapore 117542}

\date{\today}

\begin{abstract}
\noindent There have been various claims that the Equivalence Principle, as originally formulated by Einstein, presents several difficulties when extended to the quantum domain, even in the regime of weak gravity. Here we point out that by following the same approach as used for other classical principles, e.g. the principle of conservation of energy, one can, for weak fields, obtain a straightforward quantum formulation of the principle. We draw attention to a recently performed test that confirms the Equivalence Principle in this form and discuss its implications. \end{abstract}


\maketitle                           

The principle of equivalence is a pillar of Einstein's theory of relativity; as such, it was originally formulated within a classical theory, where all observables of a point particle, in particular its position, energy and mass, are sharp in any state of the particle.  
This is true of other principles, such as the principle of conservation of energy, whose expression and validity in quantum theory are nonetheless widely accepted. Yet, there has been a great degree of controversy about the formulation of the Equivalence Principle for quantum systems: this is because a quantum system can exist in a spatial superposition; and the Equivalence Principle as classically formulated does not cover directly such cases.  Consequently, there have been proposals to extend it to quantum systems, \cite{HARDY}, \cite{BRU1}, \cite{BRU2}; there also have been claims that the principle is violated by quantum systems (see e.g. the introduction in \cite{BEI}  and references herein); some have also claimed that this should be a reason for gravitational state reduction, \cite{HOPEFU}. The main point of discussion here is the fact that the Equivalence Principle implies that different masses should fall at the same rate in the same gravitational field. However, the quantum de Broglie wavelength is a function of the particle's mass and therefore different masses would interfere differently in the same gravitational field; this would seem to violate the prescription, from the Principle of Equivalence, that the behaviour of different masses in the same field cannot be distinguished. As we will see below, in our formulation of the quantum Equivalence Principle, this is not a relevant issue. The same we believe is true for other aspects of the controversy, such as those mentioned in \cite{BEI}.    

Here we would like to extend the Equivalence Principle to the quantum domain via a similar approach to that applied to the energy conservation. Namely, to extend the principle to the quantum domain, we will assume that for any branch of a quantum superposition the principle holds true. Specifically, we will assume that for each branch of a spatial superposition, sharp at location $x$, the Equivalence Principle holds in one of its currently accepted forms: {\textit {the state of motion of a point particle at rest in a uniform gravitational field ${\bf g}$ is empirically indistinguishable, by local operations at $x$, from the state of motion of a point particle that undergoes an acceleration $-{\bf g}$ in a gravity-free region}}. To the best of our knowledge, this formulation of the Equivalence Principle is original; however, there are other recent proposals that are similar in spirit \cite{HARDY}. In the limit of weak gravity this is a good approximation; the ultimate formulation of the Equivalence Principle will have to lift this assumption and it should not rely on the idea of a fixed spacetime background. For present purposes, however, it is possible to confine attention to this regime, as even in this regime problems with the quantum formulation of the Equivalence Principle have been claimed to exist. This regime is also very interesting, as it does not involve general-relativistic effects, but it can be used to probe the quantum nature of gravity, \cite{BOSE, MAVE}. We will now derive a number of well-known consequences of the principle formulated in this way and point out a recent experimental confirmation of its validity.

\noindent {\bf The phases induced by the Equivalence Principle.} Assuming the Equivalence Principle, in classical mechanics the transformation between a system in a gravitational field and the one in the equivalent accelerated 
frame is a gauge transformation. Assuming, with the Equivalence Principle, that the inertial and gravitational mass are the same constant $m$, the Lagrangian for a particle with mass $m$ moving in an accelerated frame with the acceleration rate given by $g$ is:
\[
L_F= \frac{1}{2} m({\dot x}+gt)^2 \; , 
\]
whereas for a particle moving in the gravitational potential $m_ggx$, the Lagrangian is 
\[
L_G= \frac{1}{2} m ({\dot x})^2 - mgx = L_F + \frac{d}{dt} \Lambda (x,t) 
\]
where 
\[
\Lambda = - m g x t - \frac{1}{6} mg^2 t^3
\] 
is the gauge transformation. For simplicity we only consider a one-dimensional motion, but this results in no loss of generality. 

Following the Equivalence Principle, as formulated above, the wavefunctions of the particle described in the two coordinate systems (the freely falling and the $g$ frame), as expressed in the position basis, are related in the following way \cite{GROV, NAU}:
\[
|\Psi (x,t)\rangle_G= e^{-\frac{i}{\hbar} \Lambda (x,t)} |\Psi (x,t)\rangle_F   \;.
\]
This can easily be verified by applying the Equivalence Principle in the above form to the corresponding Schr\"odinger equations: one, where the particle is freely falling; the other, where the particle is in a uniform gravitational field. The above relation holds for all the solutions of the  Schr\"odinger equation, \cite{NAU}. Now, as expected, the gauge transformation between the two coordinate systems (the freely falling and the one in the uniform gravitational field) is quantum mechanically reflected in the appearance of the extra phase factor between the two corresponding quantum states. This can be seen as a consequence of the fact that, upon quantisation, the classical Lagrangian becomes the phase factor which constitutes the basis of the path-integral formulation of quantum physics. It may at first be surprising that the gauge transformation is not just
$-mgxt$ and that there is an additional term proportional to $t^3$. 
Mathematically, it is of course clear that this term is needed since the 
freely falling Lagrangian has a $t^2$ term which, in order for the Equivalence Principle to hold, needs to be cancelled by the additional term in the gauge. 

We note in passing that General Relativity would introduce corrections to this phase. One way of thinking about it is that time simply flows at different rates at the two different heights. The difference between the two flows is given by 
\begin{eqnarray}
\Delta t & = & \left(\sqrt{1-\frac{2GM}{c^2 x}}- \sqrt{1-\frac{2GM}{c^2 (x+\Delta h)}} \right)\tau \\
& = & \left (\frac{GM}{c^2 x^2}\Delta h + \frac{1}{2} \left( \frac{GM}{c^2 x^2} \Delta h\right)^2 + ...\right)\tau
\end{eqnarray}
The phase difference can now be calculated by multiplying this by $\omega = mc^2/\hbar$ and so
\begin{equation}
\Delta \phi = mg\Delta h \tau/\hbar + mg^2 \frac{\Delta^2 h}{c^2} \tau/\hbar +...
\end{equation}

The first term in the gauge transformation was observed in the Colella-Overhauser-Werner (COW) experiment, \cite{COW}, implementing the interference of a single neutron superposed across two different heights, each experiencing different Earth gravitational fields. The potential difference between the paths $mg\Delta h$ leads to the phase difference $\Delta\phi = mg\Delta h t/\hbar$ between the two neutron states. In a variant of the COW experiment, a neutron interferometry was also performed in a uniformly accelerated interferometer, confirming the same results as the original COW experiment and thus, indirectly, the validity of the Equivalence Principle in the above form, \cite{BON}. 
The second term in this expression is the GR correction and it is clearly much smaller than the Newtonian one, therefore harder to access experimentally. As far as we can tell, it has never been observed experimentally. 

Here we want to focus on the following interesting question: is the $t^3$ term observable? We are used to hearing that gauge transformations are unobservable, so an immediate response would be ``no". However, as we will explain, it is still possible to observe the $t^3$ as a relative phase between the branches of a quantum superposition of two distinct reference frames. 

\noindent {\bf The $t^3$ phase is observable.}  To observe the $t^3$ term, we need to perform an experiment that effectively superposes the freely falling and the gravitational field states of motion for a single particle. This is the main
point of our paper. The COW experiment can certainly be analysed from the perspective of both reference frames and the two treatments ought to give the same predictions due to the Equivalence Principle. But the COW experiment
does not involve the superposition of two frames, it involves the neutron being in a superposition of different spatial paths (in Earth's frame). 

We discuss here a thought experiment where a single particle is in a superposition of two states, one where it is freely falling in a gravitational field and the other where it is static in the same gravitational field. Conceptually, this experiment could in principle be achieved as follows. 

Consider a quantum system of mass $m$ superposed across two different locations, e.g. in the state $\frac{1}{\sqrt{2}}(\ket{0}+\ket{1})\;$, and placed horizontally in an ideal, uniform gravitational field, as in the COW experiment \cite{COW}. Suppose the two branches $\ket{0}$ and $\ket{1}$ are a distance $d$ apart. As confirmed by the COW experiment, the contribution to the phase due to this field is equal on the two arms, therefore amounting to no phase difference. The interferometer can now be tilted to the vertical position (see figure 1), so that the potential on the arm labelled as $\ket{0}$ differs from that on the branch $\ket{1}$ by the amount $mgd$. At that point, the mass on say the branch $\ket{0}$ is dropped, and let interfere with the branch $\ket{1}$.

\begin{figure*}[htb]
\begin{center}
\includegraphics[scale=0.7]{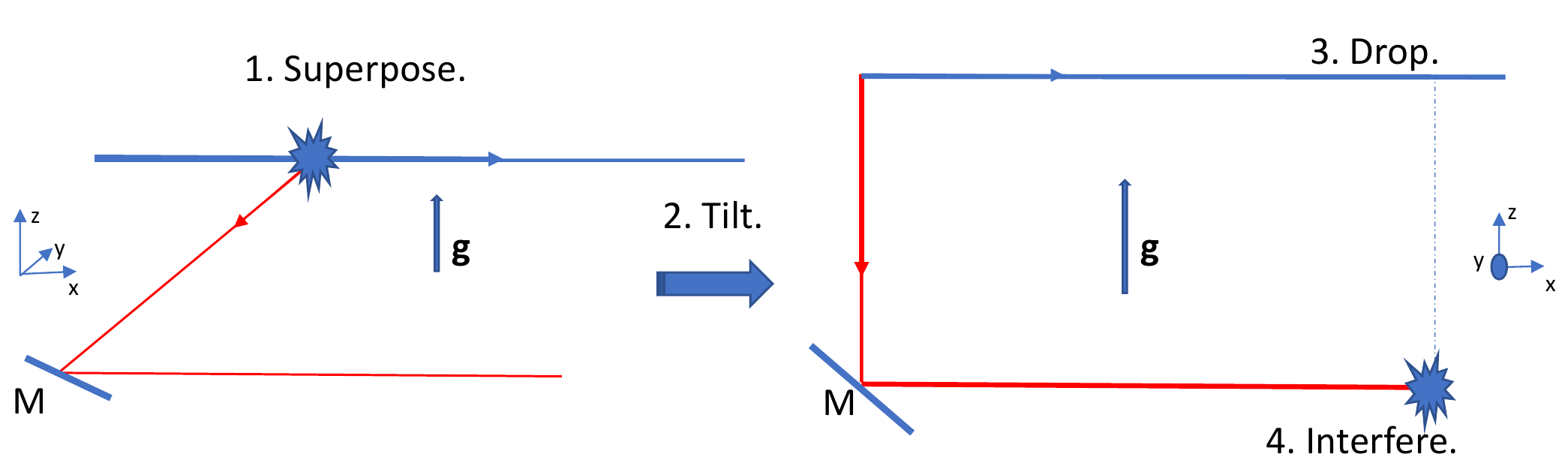}
\caption{Schematics of the interferometry to measure the $t^3$ phase. On the left, the particle is superposed across two paths (M is a mirror; {\bf g} the gravitational field). On the right, after tilting the interferometer in the vertical direction, the interferometry is closed by dropping the particle if it is on the upper branch.}
\end{center}
\end{figure*}

The phase difference between the two paths, assuming the Equivalence Principle to hold in every branch of the wave-function, and that $x(t)=d-\frac{1}{2}gt^2$, reads:

$$
\frac{1}{\hbar}\int_{0}^{t}mgx(\tau) d\tau= \frac{m g}{\hbar}  (d t  - \frac{1}{6} g t^3)
$$

The freely-falling branch of the mass interferes with the stationary branch and the resulting shift of the interference fringes contains the $t^3$ term, which can therefore in principle be detected. Whether we think of this experiment as a particle in a superposition of two different states of motion, or we think of the particle as an observer and therefore in a superposition of two reference frames, is irrelevant. The result ought to be the same, and the experiment would confirm the Equivalence Principle if the $t^3$ term were to be observed, at least in this quantum weak-field regime. 

In order to observe the phase of course no decoherence should occur. This includes any entanglement of the interfering mass to another system that does not participate in the final interference. This other system could be another simple physical system (such as a colliding atom or the electromagnetic field), or it could be an internal degree of freedom, \cite{BEI}. More subtly,
no observer should measure which of the two reference frames the mass is in, as this too would cause decoherence. 

We note that an experiment to detect the $t^3$ phase was proposed and performed very recently, with a Bose-Einstein condensate undergoing interference in a Stern-Gerlach interferometer, subject to a state-dependent force \cite{COOL}. The experiment we have discussed is conceptually equivalent. Measuring the $t^3$ in this experiment has confirmed the predictions of the Equivalence Principle expressed as above in the quantum domain. 

{\bf Discussion. } We have deployed the relation between the freely-falling and the gravitational coordinate systems, which is based on the Equivalence Principle in the weak-field limit, to discuss an experiment that could detect the $t^3$ phase that relates a mass in free fall in a gravitational field, and the same mass being stationary in the same field. We have derived this relation using a formulation of the Equivalence Principle that holds for each branch of a spatial superposition.  This expression of the Equivalence Principle harmonises with the existing ways of extending other principles, originally formulated for classical systems (e.g. energy conservation), to the quantum world. We would also like to point out that the transformation we applied to change the freely falling into the gravitational one is the weak-field limit of the Rindler coordinate transformation \cite{BIRDAV}. In this limit, it is clearly legitimate to neglect the effects such as Unruh-Davies, \cite{IVE, HOPEFU}, because only very weak fields are considered. Moreover, in the Rindler setting, the time coordinate between the inertial and the accelerated observer (with acceleration $a$) transforms as $t'=\frac{c}{a}\arcsinh \frac{a}{c} t$, where $c$ is the speed of light. Now, the second term in the Taylor expansion of this expression is equal to $\frac{a^2}{6c^2} t^3$ which, when multiplied by $mc^2$ gives us the exact $t^3$ phase term above. Therefore, the low acceleration limit of Rindler's coordinate transformation is perfectly consistent with our analysis. We conclude by pointing out that even though our discussions are in the Newtonian, weak-field quantum regime, the rest energy of the particle still somehow needs to be taken into account. This is best seen from the perspective of the relativistic action. The difference between the actions in the falling and the stationary coordinate systems can be expressed as: $mc^2(t-T)$, where $t$ and $T$ are the proper times of the two coordinate systems.  In the lowest two orders of expansion, this difference reads: $-mgxt - \frac{1}{6}mg^2t^3$. The phase difference between two branches is then, just like in the COW experiment, the difference between the time flows in those two branches (multiplied by $\omega=mc^2/\hbar$). 

Our guiding philosophy here has been to take quantum physics seriously and assume that it applies to all systems and all degrees of freedom. This means, in particular, that if any two states of motion of a particle are possible, such as an inertial and an accelerated state of a mass, then their superposition is also possible. Our paper indeed has been exploring the consequences for the relative phase between two such states, which we claim to be observable (and which the experiment in \cite{COOL} has indeed observed). 

One could also think of the measuring system being in a superposition of two different states of motion. For instance, a detector could be in a superposition of being inertial and being accelerated, while measuring another physical system in a sharp state of motion. None of this is a problem to handle quantum mechanically (as far as we can tell), but one has to be careful not to make some unwarranted assumptions. For instance, a particle being in a superposition of different motions while the detector is inertial, will in general yield different results to the detector being in different states of motion while the particle is inertial. If quantum mechanics is assumed to hold universally, none of these situations presents a difficulty: the physical systems involved will have well-defined behaviours that perfectly comply with the quantum postulates. 

In summary, given the observability of the $t^3$ term, it seems to us that at least in the regime of weak-field there should be no qualms about considering the Equivalence Principle to be extended into the quantum domain in the same way as all other classical principles are. However, there are still many interesting other open issues to be investigated both theoretically, e.g. \cite{BRU1,HARDY}, as well as experimentally.

\textit{Acknowledgments}: CM thanks the John Templeton Foundation and the Eutopia Foundation. VV's research is supported by the National Research Foundation and the 
Ministry of Education in Singapore and administered by Centre for Quantum Technologies, National University of Singapore. This publication was made possible through the support of the ID 61466 grant from the John Templeton Foundation, as part of the The Quantum Information Structure of Spacetime (QISS) Project (qiss.fr). The opinions expressed in this publication are those of the authors and do not necessarily reflect the views of the John Templeton Foundation.

\end{document}